\documentclass[osajnl,showpacs,10pt]{revtex4-2}

\usepackage{amsmath,amssymb,graphicx}
\usepackage{float}
\usepackage{multirow}
\usepackage{colortbl}
\DeclareMathOperator{\sech}{sech}
\usepackage[section]{placeins}

\begin{document}
	\title{Amplifying Optical Signals with Discrete Solitons in Waveguide Arrays}
	\author{Amaria Javed, Alaa Shaheen, and U.~Al Khawaja\\
		Department of Physics, United Arab Emirates University,\\ P.O. Box
		15551, Al-Ain, United Arab Emirates.\\ u.alkhawaja@uaeu.ac.ae}
	\date{}
	%\email{u.alkhawaja@uaeu.ac.ae}
	%\author{??????????}
	%affiliation{Department of Physics, United Arab Emirates University, P.O. Box
	%15551, Al-Ain, United Arab Emirates.}
	%\pacs{03.75.Lm; 05.45.Yv; 42.65.Tg, 42.65.Wi}
	\begin{abstract}
	We present a design and protocol to achieve an essential feature of an optical transistor, namely the amplification of input signal with the use of discrete solitons in waveguide arrays. We consider the scattering of a discrete soliton by a reflectionless potential in the presence of a control soliton. We show that at the sharp transition region between full reflectance and full transmittance, the intensity of the reflected or transmitted soliton is highly sensitive to the intensity of the control soliton. This suggests a setup of signal amplifier. For realistic purposes, we modulate the parameters of the reflectionless potential well to achieve a performance of amplifier with a controllable amplification. To facilitate the experimental realization, we calculate the amplification factor in terms of the parameters of the potential well and the input power of the control soliton. The suggested signal amplifier device will be an important component in the all-optical data processing.
\end{abstract}
	\maketitle
\section{Introduction} \label{introsec}

The growing demand for higher data processing speed and capacity motivates the replacement of the current electronic data processing by optical data processing in analogy with the successful replacement of electronic data communication by optical data communication \cite{books1, books2, books3, books4}. In a quest to achieve a comprehensive optical data processing, many of the main ingredients have been already proposed in the literature considering various setups \cite{gates1,gates2,gates3,gates4,gates5,gates6,gates7,gates8,gates9,gates10,gates11,gates12,gates13}. Prominent among these are discrete solitons in waveguide arrays \cite{gates1,gates2,gates7,gates8,gates11,gates12}. Optical solitons have been suggested as data carriers due to their unique feature of preserving their integrity over long distances of propagation and their particle-like inelastic scattering with each other and with external potentials \cite{books5,fedor1,fedor2,fedor3,usamanjp}. The optical soliton solutions of the nonlinear Schr\"odinger equation can be obtained through different methods as presented in Refs.~ \cite{Inc2,Inc3,Inc5}. The optical solitons can also be retrieved from other models like Biswas-Arshed equation \cite{Inc1}, Biswas-Milovic equation that defines the long-space optical communications \cite{Inc4} and Kundu-Eckhaus equation \cite{Inc6}. Discrete solitons are also characterised by this appealing feature. In addition, the propagation of discrete solitons in waveguide arrays can be most feasibly controlled through the dispersion or nonlinearity management \cite{andrey,lepri,usamaandrey,submitted1}.\\
\\The separations between the waveguides can be modified such that the propagating soliton experiences an effective potential. Such potential can be designed as a reflectionless potential \cite{andrey}. The management of nonlinear interaction and dispersion allows for a multitude of configurations where various data processing components can be designed. For instance, unidirectional flow has been shown to exist in waveguide arrays with dispersion management \cite{usamaandrey} and nonlinearity management \cite{submitted1} in a similar manner as for the continuum cases in optical fibers with double potential wells \cite{usamaasad}, PT-symmetric potentials \cite{usamayuri}, and nonlinearity management \cite{recentpre}.\\
\\The main objective of the present work is to achieve amplification which is an essential feature of an optical transistor in all-optical operations. The main functions of an optical transistor are switching and amplifying optical signals. The device is the optical analog of the electronic transistor that forms the basis of modern electronic devices. Optical transistors introduce the possibility of controlling light using only light and has applications in optical computing and fiber-optic communication networks. Such technology has the potential to exceed the speed of electronics, while saving more power \cite{last}. The present work is a continuation of our previous works in which switch, diode and logic gates have been proposed \cite{usamaandrey,1}, in addition to our recent design of a scheme and protocol to add binary numbers using discrete solitons in waveguide arrays \cite{submitted}.\\
\\We consider a discrete nonlinear Schr\"odinger equation with dispersion management through modulating the separations between waveguides and in the presence of a control soliton located at the minimum of the potential. We solve the equation numerically and calculate the transport coefficients from which the amplification factor is determined at the transition region between full reflection and full transmission. As the ideal reflectionless potential leads to an output power that is very sensitive to the input power, we propose lowering such unrealistic sensitivity by perturbing the parameters of the potential such that the scattering of the soliton will not be totally reflectionless. In other words, we allow for some radiation to be emitted resulting in cross-over-like type of transition rather than a sharp transition.\\
\\The rest of the paper is organized as follows. In Section \ref{modsec}, we present the setup and theoretical model. In Section \ref{figsec}, we present our main results and discussion. Finally, in Section \ref{concsec}, we summarize our main conclusions.\\\\

\section{Theoretical Model} \label{modsec}

The propagation of solitons in a one-dimensional array of $N$ waveguides with focusing nonlinearity can be described, in the tight-binding approximation, by the following discrete nonlinear Schr\"odinger equation (DNLSE) \cite{rev},
\begin{equation}\label{DNLSE}
i\dfrac{\partial \Psi_n }{\partial z}\, + C_{n,n-1} \Psi_{n-1} + C_{n,n+1} \Psi_{n+1} +
\gamma |\Psi_n|^2\Psi_n =0,
\end{equation}
where $\Psi_n$ is the normalized mode amplitude and $n$ is an
integer associated with the waveguide channel, $z$ is the
propagation distance, $C_{n,m}$ are the coupling coefficients between different
waveguide channels $n$ and $m$, and $\gamma$ is the strength of the focusing nonlinearity. Here we used $\gamma = 1$ for numerical simulation. The discrete nonlinear
Schrödinger equation is also associated with the supratransmission phenomenon which is characterized by the propagation of nonlinear localized modes (gap solitons) \cite{Motcheyo2,Motcheyo3}.\\
\\The modulation of the coupling constants through changing the separation between waveguides
leads to an effective potential and Eq.~\eqref{DNLSE} will be rewritten as \cite{andrey}:
\begin{equation}
i\dfrac{\partial \Psi_n}{\partial z} + C_{n-1}^S \Psi_{n-1} + C_{n+1}^S \Psi_{n+1} +
\gamma |\Psi_n|^2\Psi_n = 0,
\end{equation}
where
\begin{equation}\label{CM}
C_{n\pm1}^S = \sqrt{(C+|{\Psi_{n}^{AL}|^2})(C+|{\Psi_{n\pm1}^{AL}|^2)}}.
\end{equation}
For the effective potential to be reflectionless, we use the integrable Ablowitz-Ladik (AL) model
\cite{AL}
\begin{equation}\label{ALM}
i\dfrac{\partial \Psi_n}{\partial z} + (\Psi_{n-1} + \Psi_{n+1})(C+|\Psi_n|^2)=0,
\end{equation}
The above equation is one of the basic discrete equations of the integrable hierarchy. As such, it serves as
a model for the application of a range of techniques for obtaining exact solutions. The Ablowitz-Ladik equation is one of the early examples of an equation to which the inverse scattering technique has been applied. Various other methods, such as Darboux and B\"acklund transformations, have also been applied to solve this equation. Generally speaking, every technique applicable to the basic NLSE equation can also be applied, with some modifications, to the AL equation.
Here, we find a stationary bright soliton solution to the AL model, Eq.~\eqref{ALM}, using the ansatz method. Substitute the following ansatz in Eq.~\eqref{ALM}:
\begin{equation}
\psi_n=A\,{\rm sech}(\beta\,(n-n_0))\,e^{i\,\mu\,z},
\end{equation}	
where $A$, $\beta$, and $\mu$ are unknown real constants. Then, expand the resulting equation in powers of $n-n_0$. Equating coefficients of powers of $n-n_0$ to zero, gives $A=\sqrt{C}\,{\rm sinh}{\beta}$ and $\mu=2C\,{\rm cosh}{\beta}$. For these two expressions, the whole power series vanishes and thus the solution of Eq.~\eqref{ALM} will be given by
\begin{equation}\label{sol}
\Psi_{n}^{AL}=\sqrt{C}\sinh(\beta)\sech[\beta(n-n_0)]\exp(i\mu z),
\end{equation}
where $\beta$ is the inverse width of the soliton,
$n_0$ corresponds to the location of the soliton peak, and $C$ is an arbitrary real constant. A similar procedure, with the ansatz $\psi_n=A\, {\rm tanh}(\beta\,(n-n_0))\,e^{i\,\mu\,t}$, leads to the dark soliton solution.\\	
	\\All solitons used here are the stationary states of Eq.~\eqref{DNLSE}. They are generated using the Newton-Raphson method and trial solution is given by $\Psi_{n} = A \exp(-\alpha |n-n_0|)$, where $\alpha^{-1}$ and $n_0$ are parameters that set the width and peak location of the soliton respectively. As usual, two stationary modes result out of this procedure, namely, the Page mode and the Sievers-Takeno mode \cite{Boris}. Therefore, the initial condition used in our protocol can be written generally as
\begin{equation}\label{initial}
\Psi_n(0) = \Psi_s \,e^{i v n} + r~\Psi_c,
\end{equation}
where $ \Psi_s$ is the signal soliton and $\Psi_{c}$ represents input control soliton, they are both generated by the Newton-Raphson method. The coefficient $e^{i v n}$ corresponds to the kick-in velocity of the signal soliton with velocity, $v$, used to control the motion of discrete soliton. This method has the benefit of preserving the norm and it is the same kick or thrust which is used in Ref.~\cite{Motcheyo1} to control the evolution of two solitons solution in time. The parameter $r$ controls the intensity of the input control soliton $I_{in}$, related by $I_{in} = r^2$.
The intensity of the control soliton is $\sech^2(x)$-shaped and the modulation of the dispersion coefficients is also $\sech^2(x)$-shaped. The control soliton is located at the region where the dispersion is modulated. Therefore, the incident soliton will encounter two effective potentials when it is scattered. The first comes from the modulation in the dispersion coefficients and the second comes from the nonlinear interaction with the control soliton. Since the two effective potentials have the $\sech^2(x)$ shape, the net interaction energy will be determined by the relative values of the strengths of the two potentials. By controlling the intensity of the control soliton, we tune the strength of the effective potential resulting from the modulation in dispersion coefficients.
The parameter $r$ refers to an external control on the intensity of the input soliton. It can be done through tuning the intensity of the laser pulse injected into the waveguide. A more sophisticated control could be achieved by injecting the control soliton into a bean splitter where only a fraction of the soliton transmits. That fraction can be used as the input control soliton in our scheme. \\
\\The calculation of amplification factor is obtained through the transport coefficients which are defined as follows: reflection $R=\sum_{1}^{n_1-\delta n}|\Psi_n|^2 /  \sum_{1}^{N}|\Psi_n|^2$, transmission $T = \sum_{n_2+\delta n}^{N}|\Psi_n|^2 /  \sum_{1}^{N}|\Psi_n|^2$ and trapping $ L = \sum_{n_1-\delta n}^{n_2+\delta n}|\Psi_n|^2 /  \sum_{1}^{N}|\Psi_n|^2 $, where $N$ is the number of waveguides and $\delta n$ is roughly equal to the width of the soliton in order to avoid the inclusion of the tails of the trapped soliton with the reflected or transmitted ones.\\
\\In our amplification protocol, we consider the intensity of the control soliton, $I_{in}$, as the input signal and the intensity of the reflected soliton, $R$, as the output signal. Amplification is thus measured by the rate of change in the output intensity, $\Delta R$, with respect to a change in the input intensity, $\Delta I_{in}$, namely
\begin{equation}
M=\frac{\Delta R}{\Delta I_{in}}.
\end{equation}
\section{Results and Discussion} \label{figsec}
\subsection{Reflectionless Potential}

\begin{figure}[!h]
	\centering
		\includegraphics[scale=0.35]{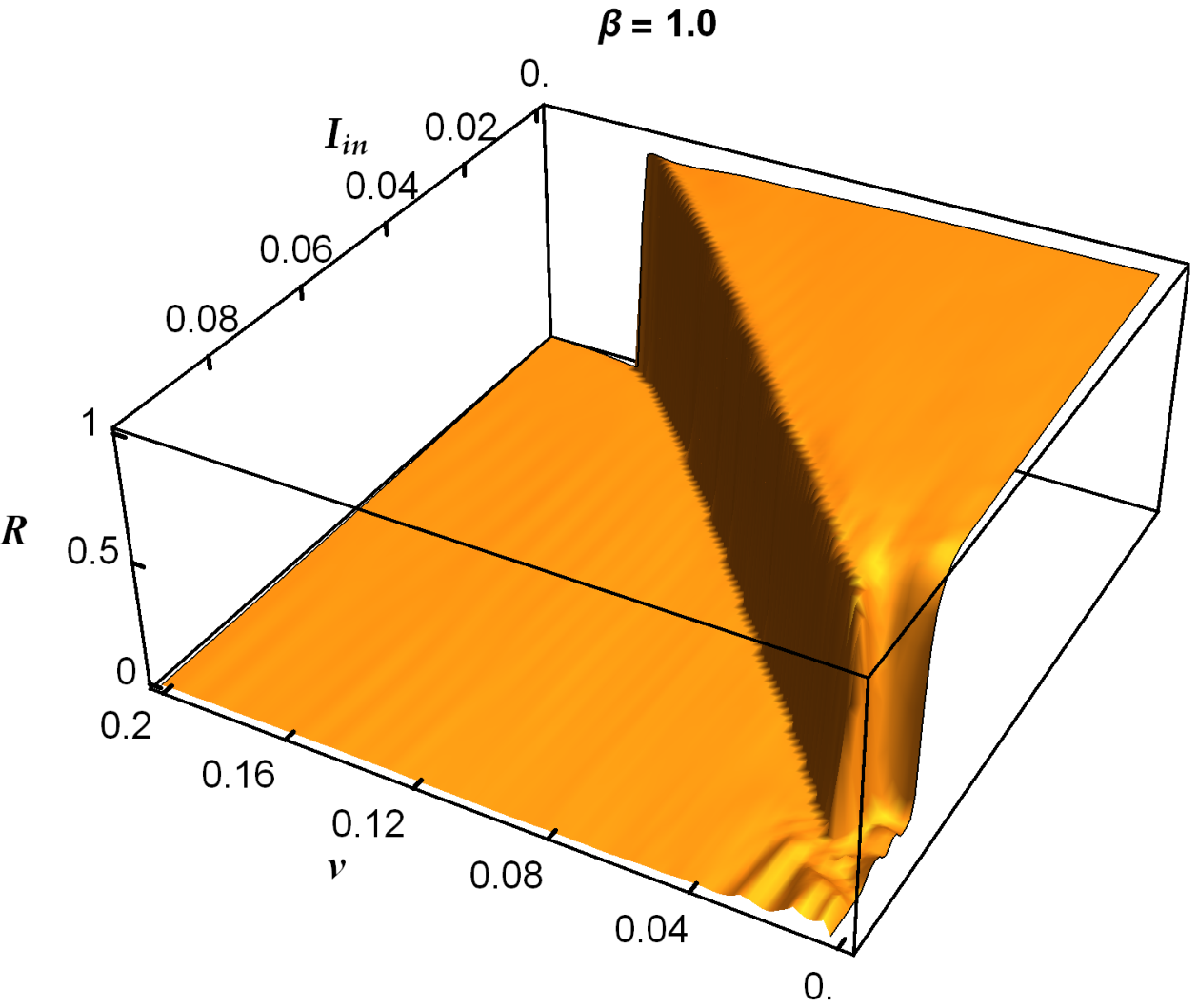}\,\,\,\,\,\,\,\,
		\includegraphics[scale=0.39]{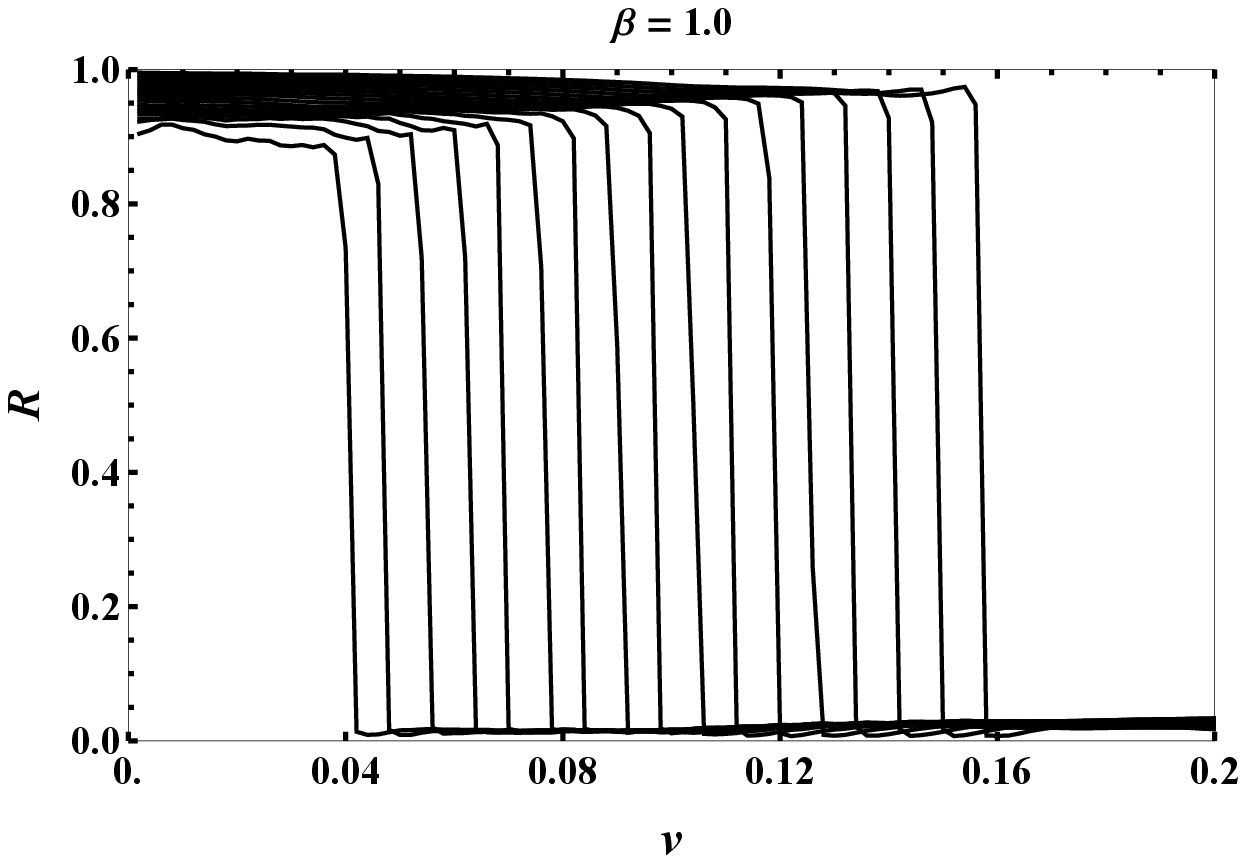}\,\,\,\,\,\,\,\,
	\includegraphics[scale=0.38]{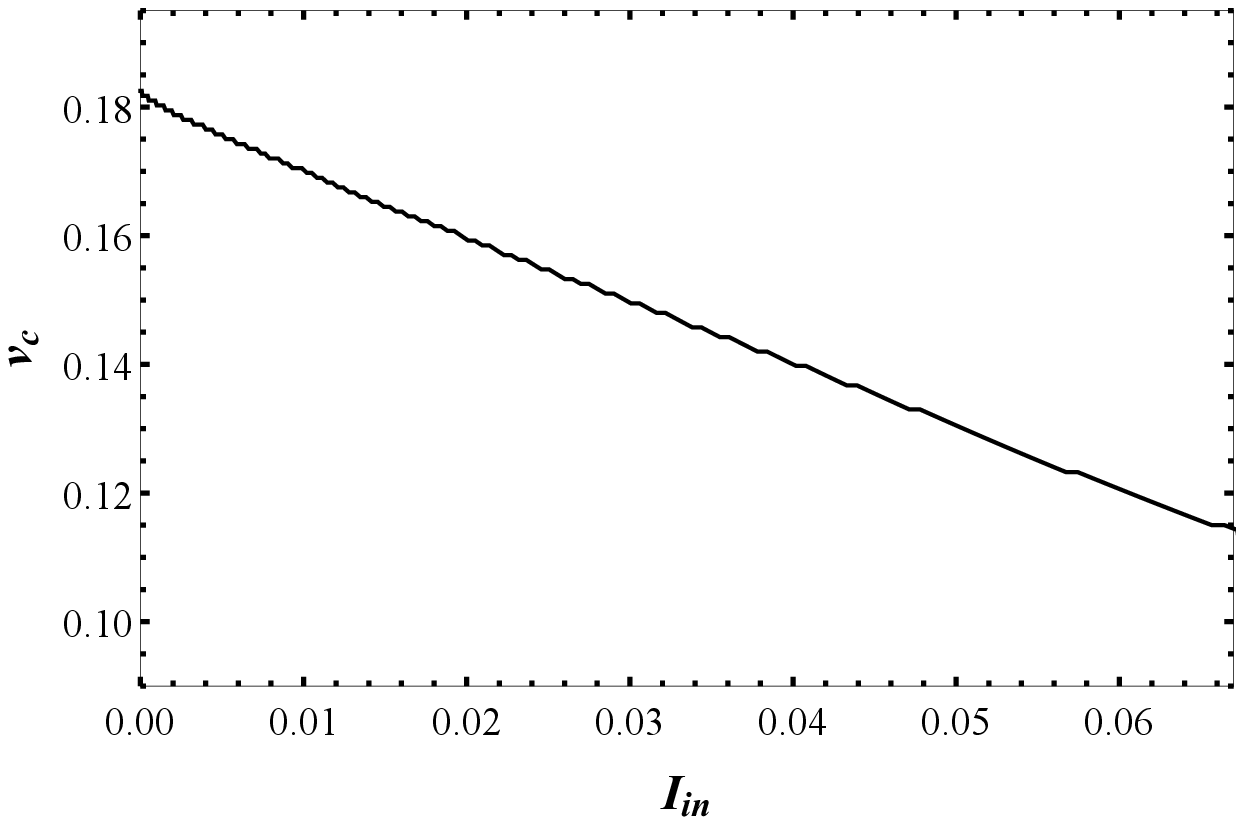}
	\caption{Reflectionless potential (Ideal case): 3D plot representing the dependence of the output intensity, $R$, on the input intensity, $I_{in}$, and signal soliton speed, $v$. The left subfigure in the lower panel corresponds to constant-$I_{in}$ cross sections from the 3D plot with $I_{in}$ reaching to $I_{in}=0.12$, for the curve with lowest critical speed, from 0.02 by an increment of 0.006. The right subfigure in the lower panel shows the critical speed in terms of the control soliton intensity. A reflectionless potential with $\beta=1$ is used.}
	\label{fig:ref-less}
\end{figure}

At first, we consider the ideal situation with a reflectionless potential. As expected and shown in Fig.~\ref{fig:ref-less}, sharp transitions occur at a critical speed, $v_c$. The intensity of the control soliton will merely shift the location at which the sharp transition occurs. \\

\noindent Since the amplification factor is proportional to the slope of the transport coefficient at the transition region, unrealistically high amplification values will be obtained. Smearing the curve is thus needed. We will achieve  this in the following by perturbing the potential such that the scattering of the soliton will not be totally reflectionless.

\subsection{Modulated Potential}
\begin{figure}[!h]
	\centering
		\includegraphics[scale=0.55]{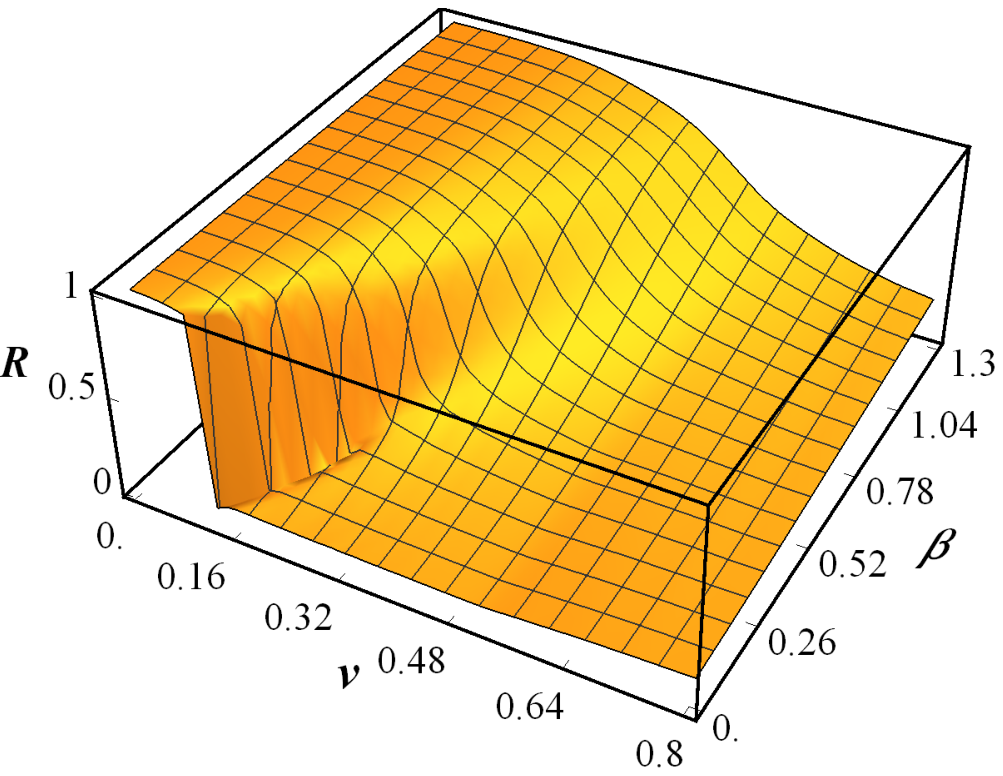}\,\,\,\,\,\,\,\,\,\,\,\,\,\,\,\,\,\,\,\,\,\,\,\,
		\includegraphics[scale=0.65]{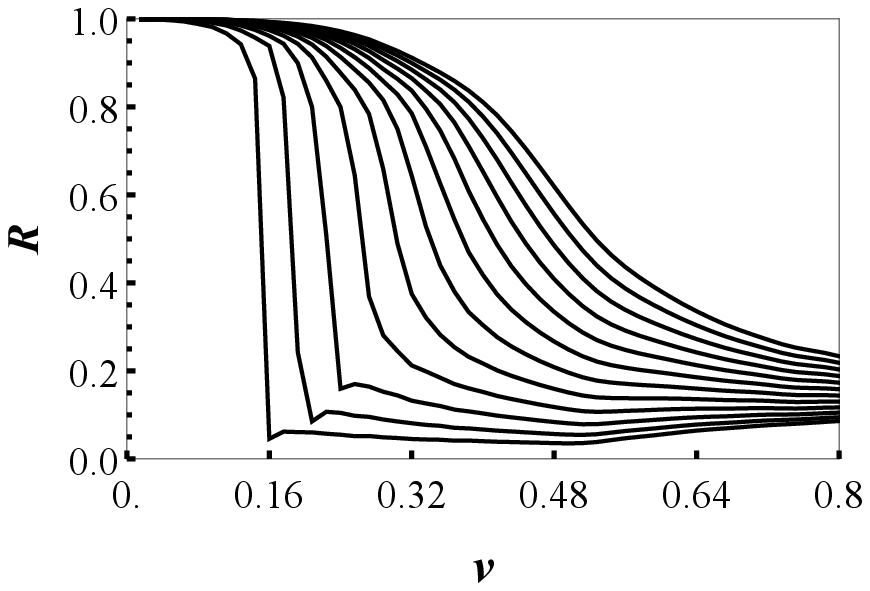}
	\caption{3D plot representing the dependence of the output intensity, $R$, on the signal soliton speed, $v$, and the parameter of the  potential well, ${\beta}$, for $I_{in}$ = 0 (no control soliton). The right subfigure shows constant-$\beta$ cross-sections.}
	\label{fig:I-in}
\end{figure}

\noindent In order to design the device with an amplification feature, we need to have a finite slope in the transition region replacing the sudden jump in the transport coefficients, as shown previously in Fig.~\ref{fig:ref-less}. This can be achieved by using a potential that is not fully reflectionless such as using
$\Psi_{n}^{AL}=\sqrt{C}\sinh(\beta)\sech[\beta(n-n_0)]\exp(i\mu z)$ with $\beta\ne1$. Using values of $\beta$ other than 1, leads to smearing the sharp transition in the transport coefficient, as shown in Fig.~\ref{fig:I-in}.

\begin{figure}[!h]
	\centering
		{\includegraphics[scale=0.45]{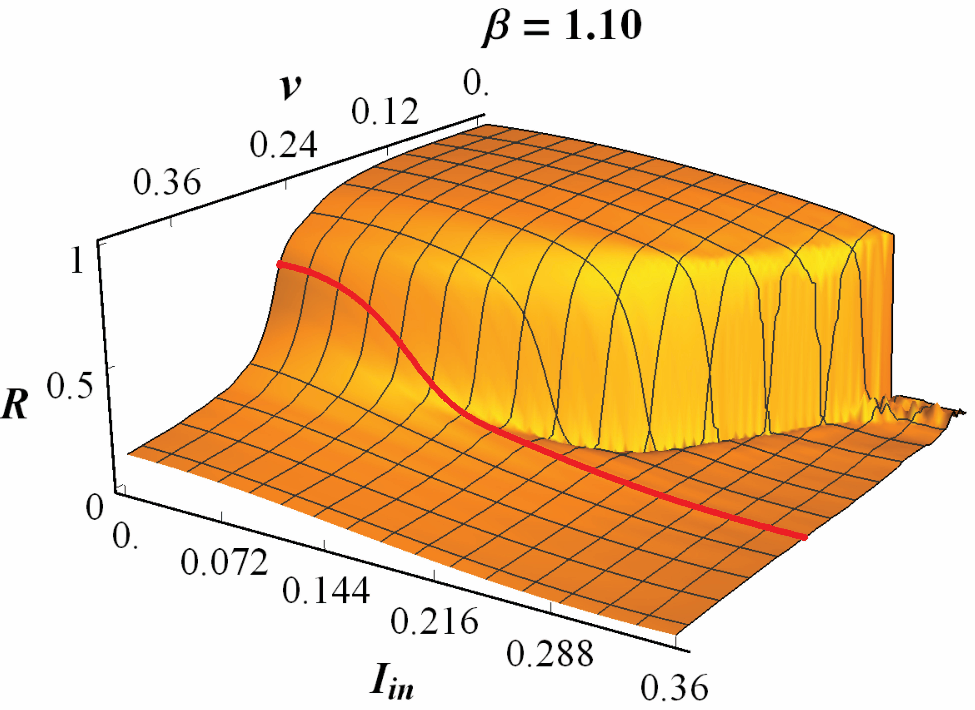}\,\,\,\,\,\,\,
		\includegraphics[scale=0.45]{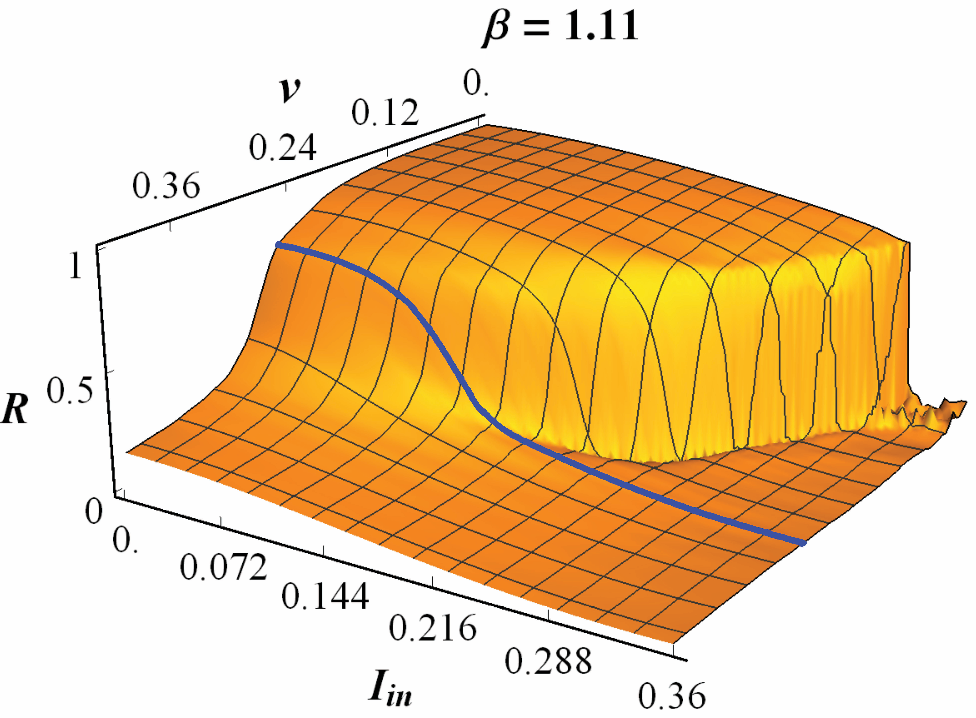}\,\,\,\,\,\,\,
		\includegraphics[scale=0.45]{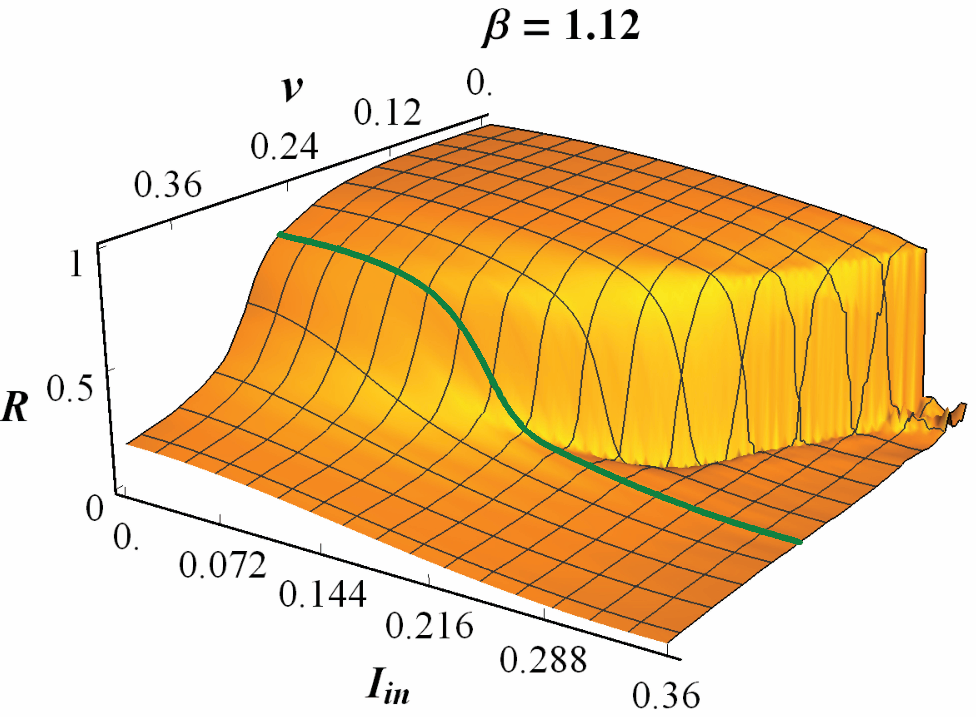}}\\
		{\includegraphics[scale=0.55]{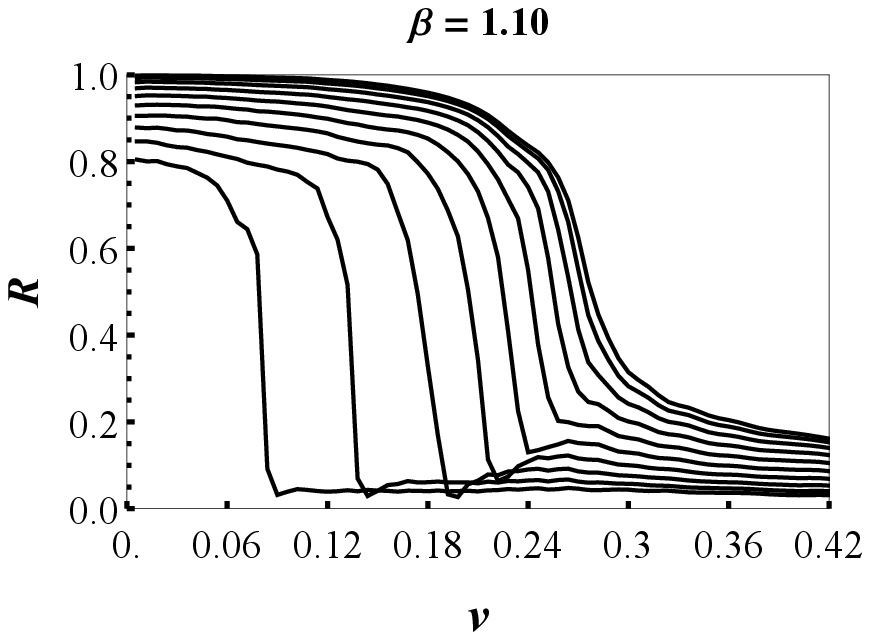}\,\,\,\,\,\,\,
		\includegraphics[scale=0.55]{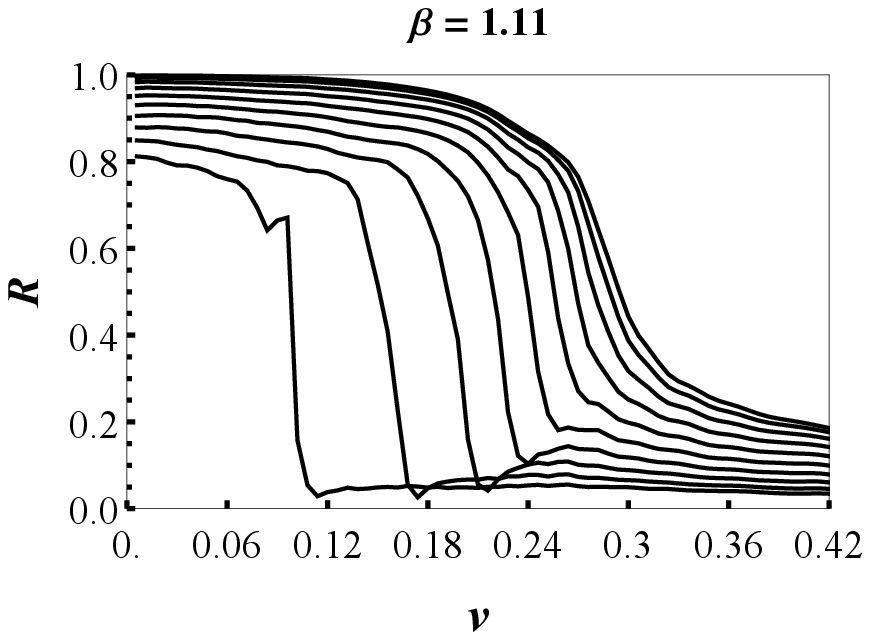}\,\,\,\,\,\,\,
		\includegraphics[scale=0.55]{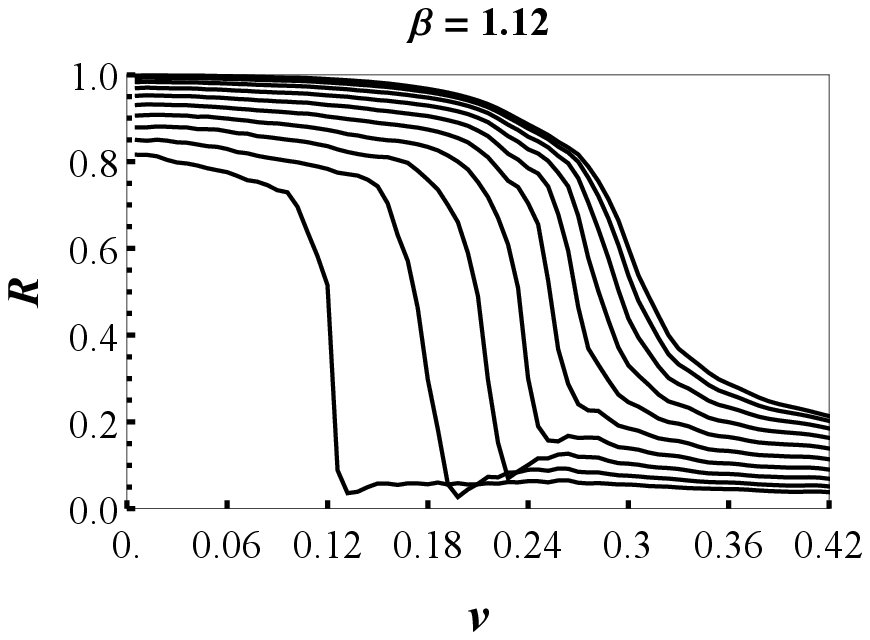}}
	\caption{Modulated potential (non-reflectionless potential): 3D plots representing the dependence of the output intensity, $R$, on the input intensity $I_{in}$, and the signal soliton speed, $v$ for three values of $\beta$. Subfigures on the right correspond to constant-$I_{in}$ cross-sections.}
	\label{fig:non-ref}
\end{figure}

\noindent To investigate in more detail the effect of modulating the potential, we monitor the changes in the intensity of the reflected soliton in terms of changes in the  intensity of control soliton and signal soliton speed for three different cases of modulated potential, as shown in Fig.~\ref{fig:non-ref}.

\begin{figure}[!h]
	\centering
		\includegraphics[scale=0.55]{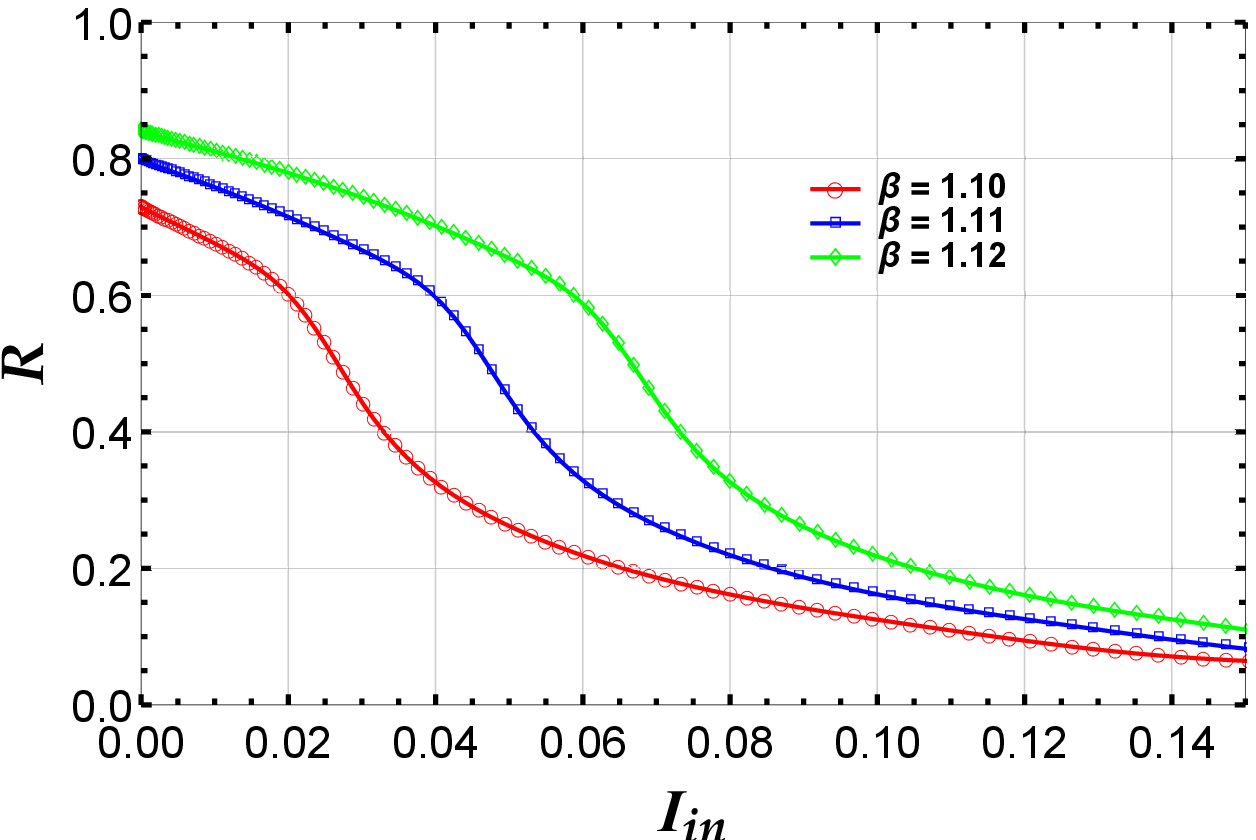}
		\includegraphics[scale=0.55]{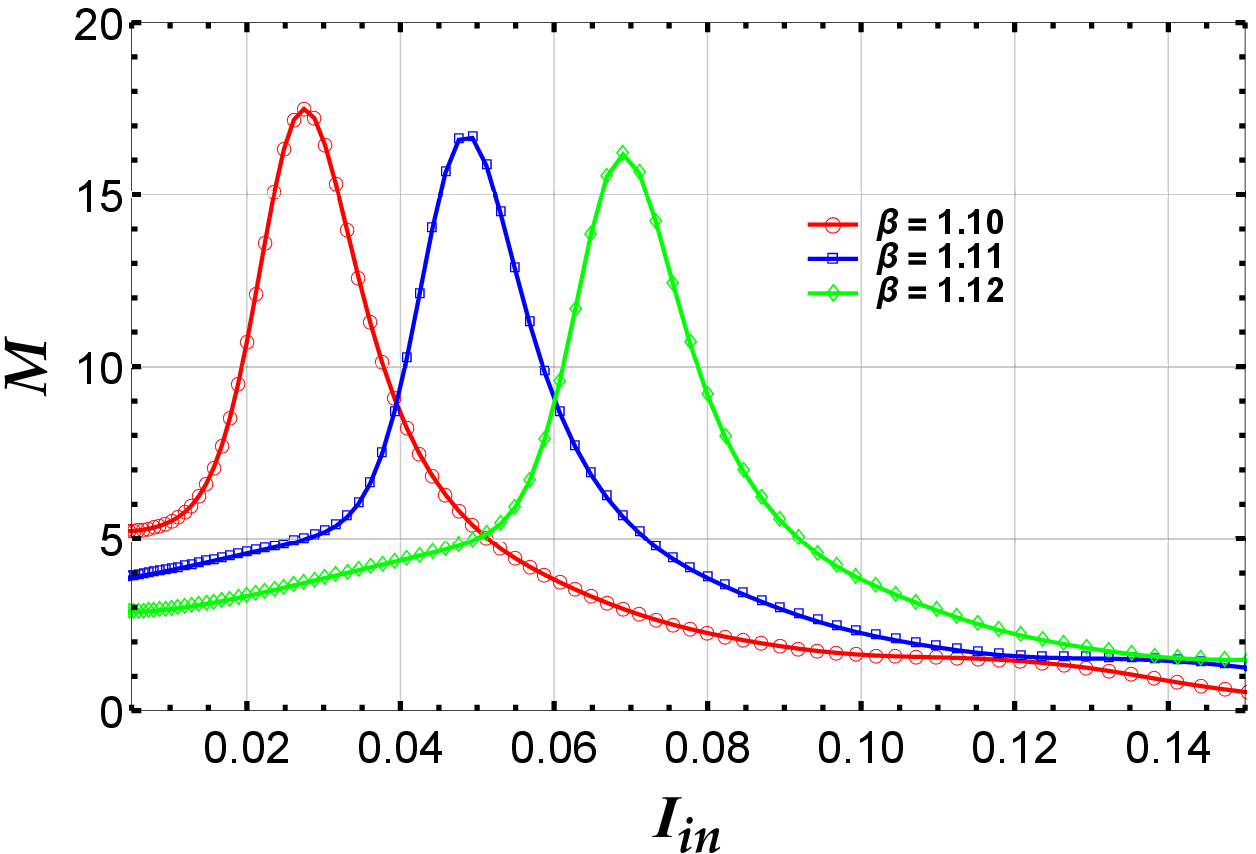}
	\caption{The left subfigure represents the relation between the output intensity, $R$, and the input intensity of control soliton, $I_{in}$, for three different cases of modified potential well $\beta$ = 1.10, $\beta$ = 1.11, and $\beta$ = 1.12. The right subfigure shows the slope of left subfigure that determines amplification. For all curves the initial soliton speed is $v$ = 0.33.}
	\label{fig:con}
\end{figure}

\noindent By fixing critical speed, $v_c$ = 0.33 in Fig.~\ref{fig:non-ref}, we investigate the dependence of intensity of reflected soliton on the intensity of control soliton for all three cases of modulated potential. The three curves in Fig.~\ref{fig:con} represent each case of modulated potential shown in Fig.~\ref{fig:non-ref} with fixed critical speed. In Fig.~\ref{fig:con}, the amplification $M$ is shown in the right subfigure corresponding to the slope of the left subfigure. The figure shows that amplification factors $M\approx17$ are achieved for the considered values of $\beta$. As mentioned previously, much larger amplification rates can be reached with values of $\beta$ approaching 1.

\section{Conclusion} \label{concsec}

We exploit the sharp transition region between full reflectance and full transmittance to achieve optical signals amplification. We find that the transition region is highly sensitive to the intensity of the input control soliton. For reflectionless potential, the sensitivity is too high to be experimentally realized. Therefore, we modulate the reflectionless potential well to achieve a realistic performance of amplifier with a controllable amplification. We also show that amplification value can be controlled by the intensity of a control soliton located at the centre of potential well and the modulation of the potential well parameters, mainly its width. We performed a detailed numerical investigation of the effect of all parameters regimes in order to optimize the performance. The separations between the waveguides can be calculated and set to achieve such a potential profile, as previously was performed for logic gates and unidirectional flow \cite{1}. We believe this to be an important and useful step towards achieving a soliton transistor and all-optical data processing.\\
\\Figure~\ref{fig:con} shows that the amplification factor is also dependent on the intensity of the input signal. This leads to nonlinear amplification, which we believe is not favorable from a practical point of view. In an ideal situation, the amplification factor aught to be constant so that no modulation of the relative amplitudes or profile of the input signal is performed while amplifying it. We consider this as a challenge for a future work where we aim at obtaining an amplification scheme with constant amplification factor within a finite range of input signal intensities.

\section*{Acknowledgment} The authors acknowledge the support of UAE University through grants UAEU-UPAR(4) 2016 and UAEU-UPAR(6) 2017.

\end{document}